\documentclass[12pt]{article}

\usepackage{comment}
\usepackage{tikz}
\usepackage{caption}

\usepackage{mathtools}

\usepackage[mathscr]{euscript}

\usepackage{slashed}

\usepackage{amsmath}

\date{}

\title{\textbf{Natural
Higher-Derivatives Generalization for the Klein-Gordon Equation
}}
\author{ \textbf{Ronaldo Thibes}
\\\\
\textit{\small{Departamento de Ci\^encias Exatas e Naturais}},\\
\textit{\small{Universidade Estadual do Sudoeste da Bahia}},\\
\textit{\small{Rodovia BR 415, km 03, s/n }},\\
\textit{\small{Itapetinga - BA, 45700-000, Brazil}}
 }

\usepackage{graphicx}
\usepackage{amssymb,amsfonts,amssymb,amsthm}
\usepackage{amsmath}

\begin{document}

\maketitle

\abstract{We propose a natural family of higher-order partial differential equations generalizing the second-order Klein-Gordon equation.  We characterize the associated model by means of a generalized action for a scalar field, containing higher-derivative terms.  The limit obtained by considering arbitrarily higher-order powers of the d'Alembertian operator leading to a formal infinite-order partial differential equation is discussed.
The general model is constructed using the exponential of the d'Alembertian differential operator.   The canonical energy-momentum tensor densities and field propagators are explicitly computed.  We consider both homogeneous and non-homogeneous situations.  The classical solutions are obtained for all cases.
}

\section*{Introduction}
Since the pioneering works of Fock, Gordon, Klein and Schr\"odinger, the Klein-Gordon (KG) equation \cite{Klein:1926tv, Gordon:1926} has established itself as one of the central and most important relations in relativistic quantum field theory being directly related to the quantum description of all elementary particles in nature.  Parallel to that fact, the KG equation has also enjoyed many different possible and practical interpretations in various branches of physics and mathematics, being successfully used both at quantum and classical levels describing from fundamental phenomena to effective theories as well as important approximating models.  In current modern notation, using metric signature convention $(+,-,\dots,-)$, the KG equation may be written as
\begin{equation}\label{KG}
(\Box+m^2)\phi=0\,,
\end{equation}
or simply $K_m\phi=0$ with
\begin{equation}\label{Km}
K_m\equiv\Box+m^2
\end{equation}
denoting what we define here as the {\it Klein-Gordon operator}.
Being (\ref{KG}) a second-order partial differential equation, it is in the essence of the mathematician or theoretical physicist to investigate its possible consistent extensions and generalizations.

The generalization of the Lagrangian framework for field theories containing higher-order derivative terms is an important however old subject, as can be seen for instance in the early references \cite{Courant-Hilbert, Chang1, Bollini:1986am}.  Actually, for mechanical systems, initial discussions of higher-derivative theories within variational principles scopes date back to \cite{Ostrogradsky, Whittaker}.  Concerning the canonical quantization process, first accounts on Hamiltonian approaches for higher-derivative field theories were presented in \cite{deWet, Coelho}.
Those initial attempts for a systematic study of higher-derivative models in quantum field theory were preceded and motivated by the introduction of the Bopp-Podolsky generalized electrodynamics based on field equations depending on fourth-order derivatives of the gauge fields \cite{Bopp, Podolsky:1942zz}.
However, since the thorough analysis by Pais and Uhlenbeck in the remarkable 1950 paper \cite{Pais:1950za},  it has become clear that the quantization of higher-derivative field theory models is not an easy task due to serious issues such as locality, causality, unitarity and negative norm states\footnote{In counterpart to field theory, in the context of ordinary quantum mechanics, higher-derivative models seem to be more amenable -- see for instance the interesting paper \cite{Nucci:2008ya}.}. 
In spite of those problems, or inspired on their circumventing challenge, the physics literature has since then experienced a blossom of related interesting and insightful papers. 
For that matter, it is practically impossible to cite all the huge amount of former generalizations of the KG equation containing higher-order derivatives present in the literature. Instead, as a representative sample, we mention here a few specific works \cite{Chiang:1975pi, Bollini:1985pk, Barci:1990in, Bollini:1992aw, Barci:1994ea, Bollini:1995iw, deUrries:1998obu}.
In reference \cite{Chiang:1975pi}, a $2n$-th-order derivative generalization for the massless Klein-Gordon equation of the form $K_0^n\phi=0$ was discussed and shown to enjoy conformal invariance.
In \cite{Bollini:1985pk}, Bollini and Giambiagi suggested that extensions of supersymmetry to higher space-time dimensions could lead to higher-derivative field equations. Along that line, the quantization of a fourth-order wave equation for a complex scalar field of the form $K_mK_{im}\phi=0$ coming from a larger supersymmetric model in six dimensions has been discussed in \cite{Barci:1990in} with a fair treatment of both bradyonic and tachyonic propagation modes. These ideas were further developed in \cite{Bollini:1992aw, Barci:1994ea} allowing for the possibility of complex $m$ in equation (\ref{KG}) above and corresponding higher-order generalizations.  A general polynomial in the d'Alembertian operator wave equation is briefly discussed in reference \cite{Barci:1994ea}, followed by a carefull treatment of a particular fourth-order case.
Building on the previous mentioned works, Bollini, Oxman and Rocca \cite{Bollini:1995iw} explored the interaction with the electromagnetic field via minimal coupling producing a family of unitary higher-order equations. Moreover, still in \cite{Bollini:1995iw}, the second-order cross section for the Compton effect in that higher-derivative context has been evaluated. 
After briefly reviewing and extending the Ostrogradsky formalism for higher-derivative field theories, Urries and Julve \cite{deUrries:1998obu} considered a similar model with arbitrary powers of the KG operator (\ref{Km}) for different masses and showed how the Ostrogradsky approach could be used to reduce the order of the corresponding field equations.

Coming to the present century, in reference \cite{Weldon:2003by}, the canonical quantization of Lorentz-invariant higher-derivative free scalar fields via Ostrogradsky reduction is performed by suppresing the spatial dependence via Fourier transform and considering only the remaining time evolution.  
A different alternative approach can be seen in \cite{Cherman:2012owa}, where a road to a general solution in $1+1$ dimensions to equations of the form $K_0^n\phi=0$ is suggested.  By extending d'Alembert well-known ingenuous method of light-front variables \cite{DAlembert:1747}, the authors of \cite{Cherman:2012owa} obtain the explicit general solution for the fourth and sixth-order massless Klein-Gordon equations with arbitrary initial conditions, considering both homogeneous and non-homogeneous cases.
In reference \cite{Kim:2013waf}, the authors explore connections with reduced-order equivalent models and perform the BRST quantization of a sixth-order derivative scalar field theory.  Another interesting path for obtaining higher-derivative models can be seen in \cite{El-Nabulsi:2016eur}, where it is shown that nonlocality in space-time via the introduction of forward and backward fields can lead to higher-order derivatives with a particular example of a fourth-order generalization of the KG equation given.
In this day and age, the study of higher-derivative theories continues to be an active lively field, still containing various chalenging problems and being the main subject of a significative number of works in the most recent literature \cite{Thibes:2016ivt, Mishra:2018hxc, Abakumova:2018eck, Nogueira:2018jdm, Manavella:2019bky, Ji:2019phv, Borges:2019gpz, Dai:2020rxm, Bonin:2019xss}.  Besides their own mathematical interest, the address of
important issues and open questions in physics such as the quantization of gravity and its alternative modified theories, the construction of effective low energy field theories, the dark energy and matter problems and the possibility of Lorentz symmetry breaking at a fundamental level in cosmological models has granted higher-derivative models an important role in the current contemporary physics scenario \cite{Barth:1983hb, Buchbinder:1992rb, Tomboulis:2015esa, Belenchia:2016bvb, Borges:2019lxm}. For all the above reasons, we understand that studies concerning new aspects of higher-derivative models are clearly more than justified. 

In the present paper, we propose a natural generalization of the KG equation leading to a consistent formal infinite-order partial differential equation.  In the sequel, we are chiefly interested in the classical analysis of the proposed model.  For that purpose, we introduce the model through a Lagrangian density, obtain its Euler-Lagrange field equation and discuss its classical solutions as a classical field theory.  The proposed generalization contains a single dimensionfull parameter $a$ associated to an intrinsic physical property of the scalar field.  In the usual second-order KG case (\ref{KG}), this parameter $a=1/m$ is directly related to the massive excitations for the corresponding quantum field, in this sense we may say that $a$ plays the role of the field generalized inverse mass.  After considering that $2n$-th order derivative model,
we take an additional natural step of extrapolating the number of terms in the characteristic sums in order to accomodate for the possibility of a Lagrangian depending on an infinite number of higher-order space-time derivatives.  The resulting 
series is then considered from a formal point of view with no claims regarding convergence issues.

For the reader's convenience, this article is organized as follows.  In section {\bf 1} below, we give a brief review of the Lagrangian framework for higher-derivative field theories introducing our notation and conventions.  Along the review, we extend the Euler-Lagrange equation as well as the expressions for the canonical energy-momentum tensor density for a general higher-derivative field theory.
In section {\bf 2}, we introduce a particular family of natural higher-derivative actions for the Klein-Gordon scalar field based on a power series constructed from the d'Alembertian differential operator.   The family neatly generalizes the usual KG action.  We discuss the limit of the series when the power series can be represented by an exponential function of the d'Alembertian operator. In section {\bf 3} we compute the energy-momentum tensor corresponding to the previous family of natural generalizations for the KG action and obtain their classical solutions.   We end the paper in section {\bf 4} with a final discussion on the propagators and on the non-homogeneous case where an external source for the Klein-Gordon field is introduced.

\section{Field Equation and Energy-Momentum Tensor for Higher-Derivative Theories}
In this first section, in order to fix our notation and conventions, we briefly review the generalizations of the Euler-Lagrange field equation and energy-momentum tensor density for Lagrangian functions depending on higher-order derivatives \cite{Courant-Hilbert, Bollini:1986am, Barut:1980aj}.  Consider a physical system described by a scalar field $\phi(x)$ living in a $D$-space-time-dimensional Minkowski space $\cal M$.  More precisely, $\phi$ denotes an $N$-times differentiable application from ${\cal D}\subset {\cal M}$ to the real line $\mathbb{R}$.  We use standard Lorentz covariant notation writting $x\equiv x^\mu\in{\cal M}$ and flat metric tensor $\eta^{\mu\nu}$ with signature diag $\eta^{\mu\nu}= (1,-1,\dots,-1)$ for $\mu,\nu=0,\dots,D-1$.  Furthermore, we assume our system to be described by a local action
\begin{equation}\label{S}
S=\int_{\cal D} d^Dx\,{\cal L}
\,,
\end{equation}
where ${\cal L}$ denotes a Lagrangian density which may depend on $\phi$ and up to its $N$-th order space-time derivatives\footnote{Although in this paper we shall be interested in models depending on a single scalar field, the generalization of the results in this section for an arbitrary number of fields $\phi^I$ is straightforward: just include the additional internal index $I$ for the fields and sum when required.}.  In other words, we have
\begin{equation}\label{L}
{\cal L}={\cal L}(\phi,\partial_{\mu_1}\phi,
\partial_{\mu_1}\partial_{\mu_2}\phi,\dots,\partial_{\mu_1}\partial_{\mu_2}\dots\partial_{\mu_N}\phi)
\,.
\end{equation}

The classical dynamics of the system is determined by demanding the action $S$ to remain stationary under arbitrary variations of $\phi$ and its derivatives, vanishing at the integration boundary $\partial{\cal D}$.
In symbols, this amounts to imposing 
\begin{equation}\label{deltaS}
\delta S = \int_{\cal D} d^Dx\,
\sum_{n=0}^{N}\frac{\partial{\cal L}}{\partial(\partial_{\mu_{(n)}}\phi)}
\delta(\partial_{\mu_{(n)}}\phi)
=0\,,
\end{equation}
with the restrictions
\begin{equation}\label{bc}
\left.\delta(\partial_{\mu_{(n)}}\phi)\right|_{\partial{\cal D}} = 0\,,\,\,\,\,\,n=0,\dots,N-1\,.
\end{equation}
In the two previous equations, we have introduced the definition\footnote{In case $n=0$, $\partial_{\mu_{(0)}}$ denotes no derivative at all, that is,
$
\partial_{\mu_{(0)}}\phi\equiv\phi
$.}
\begin{equation}
\partial_{\mu_{(n)}}\phi \equiv \partial_{\mu_1}\partial_{\mu_2}\dots\partial_{\mu_n}\phi
\,.
\end{equation}
To be more clear, the compact sum with $n$ running from $0$ to $N$ within the integrand in equation (\ref{deltaS}) above means
\begin{eqnarray}\label{deltaLo}
\sum_{n=0}^{N}\frac{\partial{\cal L}}{\partial(\partial_{\mu_{(n)}}\phi)}
\delta(\partial_{\mu_{(n)}}\phi)&=&
\frac{\partial{\cal L}}{\partial\phi}
\delta\phi
+
\frac{\partial{\cal L}}{\partial(\partial_{\mu_1}\phi)}
\delta(\partial_{\mu_1}\phi)
+
\frac{\partial{\cal L}}{\partial(\partial_{\mu_1}\partial_{\mu_2}\phi)}
\delta(\partial_{\mu_1}\partial_{\mu_2}\phi)
\nonumber\\&&
+\,\,\,\,\,\,
\dots\,\,\,\,\,\,
+
\frac{\partial{\cal L}}{\partial(\partial_{\mu_1}\partial_{\mu_2}\dots\partial_{\mu_N}\phi)}
\delta(\partial_{\mu_1}\partial_{\mu_2}\dots\partial_{\mu_N}\phi)
\end{eqnarray}
and we adopt the usual repeated index sum convention as well -- for instance the second term on the RHS of (\ref{deltaLo}) stands for
\begin{equation}
\frac{\partial{\cal L}}{\partial(\partial_{\mu_1}\phi)}
\delta(\partial_{\mu_1}\phi)\equiv
\frac{\partial{\cal L}}{\partial(\partial_{0}\phi)}
\delta(\partial_{0}\phi)
+\frac{\partial{\cal L}}{\partial(\partial_{1}\phi)}
\delta(\partial_{1}\phi)
+\,\,\,\,\dots\,\,\,\,+
\frac{\partial{\cal L}}{\partial(\partial_{D-1}\phi)}
\delta(\partial_{D-1}\phi)
\,.
\end{equation}
Note that, with this compact notation,  equation (\ref{L}) in particular could have been equivalently written simply as
\begin{equation}\tag{\ref{L}}
{\cal L}={\cal L}(\partial_{\mu_{(n)}}\phi)
\,,\,\,\,\,\,\,n=0,\dots,N\,.
\end{equation}
By using the Leibniz product derivative rule and the boundary conditions prescribed above in equations (\ref{bc}), the stationarity condition (\ref{deltaS}) can be rewritten as
\begin{equation}
\int_{\cal D}\,d^Dx
\sum_{n=0}^{N}(-1)^n
\left[
\partial_{\mu_{(n)}}
\frac{\partial{\cal L}}{\partial(\partial_{\mu_{(n)}}\phi)}
\right]
\delta\phi
=0
\,.
\end{equation}
Since the variation in the field $\delta\phi$ is arbitrary, this leads immediately to the generalized Euler-Lagrange field equation
\begin{equation}\label{ELhd}
\sum_{n=0}^{N}(-1)^n
\partial_{\mu_{(n)}}
\frac{\partial{\cal L}}{\partial(\partial_{\mu_{(n)}}\phi)}
=0\,,
\end{equation}
valid for a Lagrangian density $\cal L$ depending on up to $\phi$'s $N$-th space-time derivatives, i.e., of the form (\ref{L}). Once more for the sake of notation clarity, we note that in the particular case $N=3$, equation (\ref{ELhd}) can be written explicitly as 
\begin{equation}
\frac{\partial{\cal L}}{\partial\phi}
-\partial_\mu\frac{\partial{\cal L}}{\partial(\partial_\mu\phi)}
+\partial_\mu\partial_\nu\frac{\partial{\cal L}}{\partial(\partial_\mu\partial_\nu\phi)}
-\partial_\mu\partial_\nu\partial_\lambda\frac{\partial{\cal L}}{\partial(\partial_\mu\partial_\nu\partial_\lambda\phi)}
=0\,,
\end{equation}
where we have redefined the indexes
$\mu\equiv\mu_1$, $\nu\equiv\mu_2$ and $\lambda\equiv\mu_3$, reproducing thus the familiar result for the alternating sign generalized third-order Euler-Lagrange field equation.

Proceeding next to the energy-momentum tensor, we see that corresponding to the space-time translational invariance of action (\ref{S}), we have Noether conserved currents $T^{\alpha\mu}$ satisfying
\begin{equation}
\partial_\alpha T^{\alpha\mu}=0
\,.
\end{equation}
The quantities $T^{\alpha\mu}$ denote the higher-order generalization of the canonical energy-momentum tensor density given by
\begin{equation}\label{T}
T^{\alpha\mu}=
-\eta^{\alpha\mu}{\cal L}+
\sum_{n=0}^{N-1}\sum_{m=0}^{N-n-1}(-1)^m
\partial_{\mu_{(m)}}
\frac{\partial{\cal L}}
{\partial(\partial_\alpha\partial_{\mu_{(m)}}\partial_{\nu_{(n)}}\phi)}
\partial_{\nu_{(n)}}\partial^\mu\phi
\,,
\end{equation}
with the natural number $N$, already defined in equation (\ref{L}), denoting the maximum order of $\phi$ space-time derivatives appearing in $\cal L$.
Alternatively, by regrouping terms, we may cleverly rewrite (\ref{T}) also as
\begin{equation}\label{Talt}
T^{\alpha\mu}=
-\eta^{\alpha\mu}{\cal L}+
\sum_{n=0}^{N-1}\sum_{m=0}^n
\partial_{\mu_{(m)}}
\frac{\partial{\cal L}}
{\partial(\partial_\alpha\partial_{\mu_{(m)}}\partial_{\nu_{(n-m)}}\phi)}
\partial_{\nu_{(n-m)}}\partial^\mu\phi
\,,
\end{equation}
where now each value of $n$ in the external sum of the second term in the RHS is related to the Lagrangian density derivative with respect to $\phi$'s $(n+1)$-order space-time derivative.  This second form (\ref{Talt}) turns out to be the most convenient for our practical purposes when applied to specific higher-derivative models, as will be seen in the next sections.
In the particular case $N=3$, the Lagrangian density (\ref{L}) depends on up to $\phi$'s third space-time derivative and the nested sums for $T^{\alpha\mu}$ in (\ref{T}) can be explicitly worked out as 
\begin{eqnarray}\label{T3}
T^{\alpha\mu}&=&
\left[
\frac{\partial{\cal L}}{\partial(\partial_\alpha\phi)}
-\partial_{\mu_1}\frac{\partial{\cal L}}{\partial(\partial_\alpha\partial_{\mu_1}\phi)}
+\partial_{\mu_1}\partial_{\mu_2}\frac{\partial{\cal L}}{\partial(\partial_\alpha\partial_{\mu_1}\partial_{\mu_2}\phi)}
\right]\partial^\mu\phi
\nonumber\\
&&+
\left[
\frac{\partial{\cal L}}{\partial(\partial_\alpha\partial_{\nu_1}\phi)}
-\partial_{\mu_1}\frac{\partial{\cal L}}{\partial(\partial_\alpha\partial_{\mu_1}\partial_{\nu_1}\phi)}
\right]\partial_{\nu_1}\partial^\mu\phi
\nonumber\\
&&+
\left[
\frac{\partial{\cal L}}{\partial(\partial_\alpha\partial_{\nu_1}\partial_{\nu_2}\phi)}
\right]\partial_{\nu_1}\partial_{\nu_2}\partial^\mu\phi
-\eta^{\alpha\mu}{\cal L}
\,,
\end{eqnarray}
while (\ref{Talt}) leads to
\begin{eqnarray}
T^{\alpha\mu}&=&\frac{\partial{\cal L}}{\partial(\partial_\alpha\phi)}\partial^\mu\phi
+\left[
\frac{\partial{\cal L}}{\partial(\partial_\alpha\partial_{\nu_1}\phi)}\partial_{\nu_1}\partial^\mu\phi
-
\partial_{\mu_1}\frac{\partial{\cal L}}{\partial(\partial_\alpha\partial_{\mu_1}\phi)}\partial^\mu\phi
\right]
\nonumber\\&&
+\left[
\frac{\partial{\cal L}}{\partial(\partial_\alpha\partial_{\nu_1}\partial_{\nu_2}\phi)}\partial_{\nu_1}\partial_{\nu_2}\partial^\mu\phi
-
\partial_{\mu_1}\frac{\partial{\cal L}}{\partial(\partial_\alpha\partial_{\mu_1}\partial_{\nu_1}\phi)}\partial_{\nu_1}\partial^\mu\phi
\right.\nonumber\\&&\left.
+
\partial_{\mu_1}\partial_{\mu_2}
\frac{\partial{\cal L}}{\partial(\partial_\alpha\partial_{\mu_1}\partial_{\mu_2}\phi)}\partial^\mu\phi
\right]
-\eta^{\alpha\mu}{\cal L}
\,,
\end{eqnarray}
which is of course the same as (\ref{T3}).
It is easy to see the regularity pattern in the two equations above, clarifying the sums in the general expressions (\ref{T}) and  (\ref{Talt}) valid for arbitrary $N$.
In the next section, we propose a specific natural higher-derivative scalar field Lagrangian generalization to the well-known second-order Klein-Gordon case.

\section{A Family of Natural Higher-Derivatives Actions for the Klein-Gordon Field}
The d'Alembertian operator $\Box\equiv\partial_\mu\partial^\mu$ is a regular local covariant second-order differential operator, more than suitable to be recursively and consistently applied to the scalar field $\phi(x)$.
Hence, we look for a generalization for the Klein-Gordon Lagrangian containing natural powers of $\Box$.  To turn it dimensionless, we may introduce a length-dimensional constant multiplicative factor $a>0$. In terms of power series, the {\it exponential function} characterizes one of its main representatives.
Thus, by considering the individual Lagrangian density pieces\footnote{A study of the isolated pieces ${\cal L}_n$, each characterizing an independent system on its own can be found for instance in reference \cite{Chiang:1975pi}.}
\begin{equation}\label{Ln}
{\cal L}_n\equiv-\frac{a^{2(n-1)}}{2n!}\phi\Box^n\phi
\end{equation}
for $n\in\mathbb{N}$,
we define the total $2N$-th order Lagrangian density
\begin{equation}\label{L2N}
{\cal L}^{(2N)}\equiv\sum_{n=0}^{n=N}{\cal L}_n
=-\frac{1}{2}\phi\left(\sum_{n=0}^N\frac{a^{2(n-1)}\Box^n}{n!}\right)\phi
\,.
\end{equation}
It is then natural to investigate the behavior of (\ref{L2N}) in the limit of arbitrarily large $N$, for which we define further the complete Lagrangian density
\begin{equation}\label{Lphi}
{\cal L}_\phi\equiv-\frac{1}{2a^2}\phi e^{a^2\Box}\phi
\end{equation}
in terms of the d'Alembertian exponential differential operator $e^{a^2\Box}$.
Performing a space-time integration,
corresponding to the Lagrangian densities (\ref{L2N}) and (\ref{Lphi}), we define respectively the actions
\begin{equation}\label{S2N}
S^{(2N)}= -\frac{1}{2}\int d^Dx\,\phi(x)
\sum_{n=0}^N a^{2(n-1)}\frac{\Box^n\phi(x)}{n!}
\end{equation}
and
\begin{equation}\label{Sphi}
S_\phi =-\frac{1}{2a^2}\int d^Dx \phi(x) e^{a^2\Box}\phi(x)
\,.
\end{equation}

Substituting (\ref{L2N}) into the general expression (\ref{ELhd}), 
the Euler-Lagrange field equation associated to ${\cal L}^{(2N)}$ can be readily computed as
\begin{equation}\label{EL2N}
\sum_{n=0}^N\frac{a^{2(n-1)}\Box^n}{n!}\phi=0
\,,
\end{equation}
representing thus a $2N$-th-order homogeneous partial differential equation for the unknown field $\phi$.
Note that (\ref{EL2N}) reproduces the customary second-order Klein-Gordon equation
\begin{equation}
(\Box+\frac{1}{a^2})\phi=0
\end{equation}
in the particular case $N=1$.  By direct comparison with (\ref{KG}), we see that $1/a$ corresponds to the field excitation mass.  For the next particular value $N=2$, corresponding to ${\cal L}^{(4)}$ in equation (\ref{L2N}), the three-terms-sum resulting from (\ref{EL2N}) generates the fourth-order generalized Klein-Gordon equation
\begin{equation}\label{N=2}
\Box^2\phi=-\frac{2}{a^2}(\Box+\frac{1}{a^2})\phi
\,.
\end{equation}
In the limit of arbitralily large $N$, the Euler-Lagrange equation (\ref{EL2N}) can be symbolically written as
\begin{equation}\label{ea2B}
e^{a^2\Box}\phi = 0
\,.
\end{equation}
Naturally, relation (\ref{ea2B}) can be interpreted as a formal infinite-order partial differential equation, precisely the Euler-Lagrange equation resulting from the application of the stationarity variational principle to action (\ref{Sphi}).  In the next section, we obtain the energy-momentum tensor and classical solutions for these generalizations of the Klein-Gordon action.

\section{Energy-Momentum Tensor and Classical Solutions}
In this section we obtain the canonical energy-momentum density and discuss the classical solutions associated to the previous 
generalized higher-derivative actions.
Considering that the generalized KG action (\ref{S2N}) is constructed by assembling together the elementary terms ${\cal L}_n$ defined by (\ref{Ln}), from equation (\ref{Talt}) we can immediately write the energy-momentum tensor density corresponding to $S^{(2N)}$ as
\begin{equation}\label{T2Ninit}
\overset{\scriptscriptstyle\!\!\!\!(2N)}{T^{\alpha\mu}}
=
\sum_{n=0}^N
\left\lbrace
-\eta^{\alpha\mu}{\cal L}_n
+
\sum_{m=0}^{2n-1}{(-1)}^m
\partial_{\mu_{(m)}}\frac{\partial{\cal L}_n}{\partial(\partial_\alpha\partial_{\mu_{(m)}}\partial_{\nu_{(2n-m-1)}}\phi)}
\partial_{\nu_{(2n-m-1)}}\partial^\mu\phi
\right\rbrace
\,.
\end{equation}
It is interesting, and indeed useful, to observe that this last equation
(\ref{T2Ninit}) can be rewritten recursively as
\begin{equation}\label{T2Nrec}
\overset{\scriptscriptstyle\!\!\!\!(2N)}{T^{\alpha\mu}}
=\,
\overset{\scriptscriptstyle\!\!\!\!(2N-2)}{T^{\alpha\mu}}
-\eta^{\alpha\mu}{\cal L}_N
+
\sum_{m=0}^{2N-1}{(-1)}^m
\partial_{\mu_{(m)}}\frac{\partial{\cal L}_N}{\partial(\partial_\alpha\partial_{\mu_{(m)}}\partial_{\nu_{(2N-m-1)}}\phi)}
\partial_{\nu_{(2N-m-1)}}\partial^\mu\phi
\,.
\end{equation}
In order to easily compute the partial derivative of ${\cal L}_N$ inside the summation in the RHS of (\ref{T2Nrec}) above, we introduce one more piece of compact notation,
\begin{equation}
\eta^{\lambda_{(2n)}}=\eta^{\lambda_1\lambda_2\dots\lambda_{2n}}\equiv \eta^{\lambda_1\lambda_2}\eta^{\lambda_3\lambda_4}\dots\eta^{\lambda_{(2n-1)}\lambda_{(2n)}}
\,,
\end{equation}
and rewrite the d'Alembertian $n$-th power in ${\cal L}_n$ as
\begin{equation}
\Box^n=\eta^{\lambda_{(2n)}} \partial_{\lambda_{(2n)}}
\,.
\end{equation}
In this way we can see immediately that
\begin{equation}
\frac{\partial(\Box^n\phi)}{\partial(\partial_{\mu_{(2n)}}\phi)}
=\eta^{<\mu_{(2n)}>}
\,,
\end{equation}
and thus
\begin{equation}\label{partialLn}
\frac{\partial{\cal L}_n}{\partial(\partial_\alpha\partial_{\mu_{(m)}}\partial_{\nu_{(2n-m-1)}}\phi)}
=
-\frac{a^{2(n-1)}}{2n!}\phi\,\eta^{<\alpha\mu_{(m)}\nu_{(2n-m-a)}>}
\,.
\end{equation}
The bracket symbol $<\dots>$ in the last two equations denotes total symmetrization in its internal indexes\footnote{For instance $\eta^{<\mu_{(4)}>}=\frac{1}{3}
\left(
\eta^{\mu_{1}\mu_{2}}\eta^{\mu_{3}\mu_{4}}+
\eta^{\mu_{1}\mu_{3}}\eta^{\mu_{4}\mu_{2}}+
\eta^{\mu_{1}\mu_{4}}\eta^{\mu_{2}\mu_{3}}
\right)$ etc.}.
Substituting (\ref{Ln}) and (\ref{partialLn}) back into (\ref{T2Nrec}), with $n=N$, we obtain the final form
\begin{eqnarray}\label{T2N}
\overset{\scriptscriptstyle\!\!\!\!(2N)}{T^{\alpha\mu}}
&=&\,
\overset{\scriptscriptstyle\!\!\!\!(2N-2)}{T^{\alpha\mu}}
-\eta^{\alpha\mu}\frac{a^{2(N-1)}}{2N!}\phi\Box^N\phi
\nonumber\\&&
+\frac{a^{2(N-1)}}{2N!}\sum_{m=0}^{2N-1}(-1)^m\eta^{<\alpha\mu_{(m)}\nu_{(2N-m-1)}>}
\partial_{\mu_{(m)}}\phi\partial_{\nu_{(2N-m-1)}}\partial^\mu\phi
\end{eqnarray}
for the canonical energy-momentum density associated to the action (\ref{S2N}).  Just for a check, if we plug the values $N=1$ and $N=2$ into expression (\ref{T2N}) we obtain
\begin{equation}
\overset{\scriptscriptstyle\!\!\!\!(2)}{T^{\alpha\mu}}
= \frac{1}{2a^2}\eta^{\alpha\mu}\phi^2+\frac{1}{2}\eta^{\alpha\mu}\phi\Box\phi
-\frac{1}{2}\phi\partial^\alpha\partial^\mu\phi+\frac{1}{2}\partial^\alpha\phi\partial^\mu\phi
\,,
\end{equation}
and
\begin{eqnarray}
\overset{\scriptscriptstyle\!\!\!\!(4)}{T^{\alpha\mu}}
&=&
\overset{\scriptscriptstyle\!\!\!\!(2)}{T^{\alpha\mu}}
+\frac{a^2}{4}\eta^{\alpha\mu}\phi\Box^2\phi
-\frac{a^2}{4}\phi\partial^\alpha\Box\partial^\mu\phi
+\frac{a^2}{12}\partial^\alpha\phi\Box\partial^\mu\phi
+\frac{a^2}{6}\partial_\lambda\phi\partial^\alpha\partial^\lambda\phi\partial^\mu\phi
\nonumber\\&&
-\frac{a^2}{6}\partial^\alpha\partial_\lambda\phi\partial^\lambda\partial^\mu\phi
-\frac{a^2}{12}\Box\phi\partial^\alpha\partial^\mu\phi
+\frac{a^2}{4}\partial^\alpha\Box\phi\partial^\mu\phi
\,,
\end{eqnarray}
which represent  respectively the energy-momentum densities for the usual KG and generalized fourth-order KG actions.

Our next goal is to investigate the classical solutions of the $2N$-th-order partial differential equation (\ref{EL2N}) for arbitrary natural $N$.  For this purpose, we start by decomposing $\phi(x)$ into Fourier modes $\tilde{\phi}(p)$ defined from
\begin{equation}\label{FT}
\phi(x) = \int \frac{d^Dp}{(2\pi)^D}
e^{-ip\cdot x}\tilde{\phi}(p)
\end{equation}
with $p\cdot x\equiv p_\mu x^\mu$ in the exponential argument.  Then, using $\partial_\nu(p_\mu x^\mu) = p_\nu$ and applying $n$ times the d'Alembertian operator to both sides of (\ref{FT}), we have
\begin{equation}
\Box^n\phi(x)={(-1)}^n\int \frac{d^Dp}{{(2\pi)}^D}
p^{2n}e^{-ip\cdot x}\tilde{\phi}(p)
\,.
\end{equation}
It is then clear that the expansion (\ref{FT}) represents a solution to (\ref{EL2N}) as long as $p_\mu$ satisfies
\begin{equation}\label{AR}
\sum_{n=0}^N
\frac{{(-1)}^n a^{2(n-1)}}{n!}p^{(2n)} = 0
\,.
\end{equation}
That means we can formally write the general solution to (\ref{EL2N}) by means of a Dirac delta distribution as
\begin{equation}\label{fgs}
\phi_{(N)}(x) = \int \frac{d^Dp}{(2\pi)^D}
\,\,\delta
\left(
\sum_{n=0}^N
\frac{{(-1)}^n {(a^{2}p^2)}^n}{n!}
\right)\,
e^{-ip\cdot x}\tilde{\phi}(p)
\,.
\end{equation}
As usual, the open functions $\tilde{\phi}(p)$ can be used to fine tune this general solution above to a particular one satisfying specific given boundary conditions for (\ref{EL2N}).

Now comes a crucial point.  What are the solutions to the algebraic relation (\ref{AR})? Note that $p_\mu$ represents a set of $D$ {\it real} variables coming from the Fourier transformation (\ref{FT}).  Thus it must be the case that $p^2\in\mathbb{R}$.  Since the Minkowski metric is not positive-definite, $p^2$ can be any real number, either positive, negative or null\footnote{As is well-known, upon quantization of the model, positive values of $p^2$ come to correspond to bradyonic particle excitations, negative values to tachyonic excitations while $p^2=0$ leads to massless excitations.  See for instance references \cite{Tanaka:1960, Feinberg:1967zza, Barci:1995cj, Peskin:1995ev, Das:2008zze}.}.  Actually, the multiple integration in equation (\ref{fgs}) above runs through all real values of $p_\mu$, while the internal Dirac delta distribution is responsible for restricting the effective non-null contributions to those combinations leading to $p^2$ satisfying (\ref{AR}).  By performing the change of variables
\begin{equation}
q\equiv a^2p^2 \in \mathbb{R}
\,,
\end{equation}
we see that, after multiplying by $N!$, equation (\ref{AR}) can be rewritten
as
\begin{equation}\label{f=0}
f_N(q)=0
\end{equation}
where $f_N(q)$ represents an $N$-th order polynomial in the dimensionless real variable $q$ defined by
\begin{equation}
f_N(q)\equiv
\sum_{n=0}^N\frac{{(-1)}^nN!}{(N-n)!}q^{N-n}
\,.
\end{equation}
It is clear that the polynomial equation (\ref{f=0}) has real solutions only for odd values of $N$.  Actually, for odd $N$, equation (\ref{f=0}) has exactly one positive real solution which we denote here by $q_N$, while for even $N$, the algebraic relation (\ref{f=0}) has no real solutions at all.  As a consequence, if $N$ is even, the homogeneous partial differential equation (\ref{EL2N}) possesses only the trivial solution $\phi\equiv0$.

On the other hand, if $N$ is an odd natural number, using the identity $f'_N(q)=N!f_{(N-1)}(q)$, we may rewrite the Dirac delta distribution within (\ref{fgs}) as
\begin{equation}
\delta
\left(
\sum_{n=0}^N
\frac{{(-1)}^n {(a^{2}p^2)}^n}{n!}
\right)
=N!\delta(f_N(q))=\frac{1}{f_{(N-1)}(q_N)}\delta(q-q_N)
\,,
\end{equation}
redefine the Fourier modes as 
\begin{equation}
\tilde\phi(p)\equiv a^2f_{(N-1)}(q_N)\bar\phi(p)
\,,
\end{equation}
and thus rewrite
\begin{equation}\label{fgs2}
\phi_{(N)}(x) = \int \frac{d^Dp}{(2\pi)^D}
\,\,\delta
\left(
p^2-q_{N}/a^2
\right)\,
e^{-ip\cdot x}\bar{\phi}(p)
\,.
\end{equation}  
Then, by defining
\begin{equation}\label{DR}
E_N(\mathbf{p}^2)\equiv\sqrt{\mathbf{p}^2-q_N/a^2}
\end{equation}
we come to a point from which
it is possible to follow the same well-known steps for the ordinary second-order Klein-Gordon equation case \cite{Peskin:1995ev, Das:2008zze, Srednicki:2007qs} and finally obtain
\begin{equation}\label{gs}
\phi_{(N)}(x) = \int \frac{d^{D-1}\mathbf{p}}{E_N(\mathbf{p}^2)}
\big\lbrace
\varphi_N(\mathbf{p})e^{-i(E_N(\mathbf{p}^2)\,t-\mathbf{p}\cdot \mathbf{x})}
+
\varphi^*_N(\mathbf{p})e^{i(E_N(\mathbf{p}^2)\,t-\mathbf{p}\cdot \mathbf{x})}
\big\rbrace\,.
\end{equation}
Hence we have obtained the general solution for the $2N$-th-order homogeneous partial differential equation (\ref{EL2N}), namely $\phi\equiv0$ for even values of $N$ and equation (\ref{gs}) above for odd values of $N$.  In the next section we investigate the non-homogeneous case. 

\section{The Non-Homogeneous Partial Differential Equations and Generalized Field Propagator}
As we have seen, in case of $N$ even, the partial differential equation (\ref{EL2N}) does not possess nontrivial solutions.  Two representative explicit cases can be written as
\begin{equation}\label{N=4}
\Big(\frac{a^6}{4!}\Box^4+\frac{a^4}{3!}\Box^3+\frac{a^2}{2}\Box^2+\Box+\frac{1}{a^2}\Big)\phi=0\,,
\end{equation}
and
\begin{equation}\label{N=6}
\Big(\frac{a^{10}}{6!}\Box^6+\frac{a^8}{5!}\Box^5+\frac{a^6}{4!}\Box^4+\frac{a^4}{3!}\Box^3+\frac{a^2}{2}\Box^2+\Box+\frac{1}{a^2}\Big)\phi=0\,,
\end{equation}
for $N=4$ and $N=6$ respectively.  Furthermore, the formal infinite-order partial differential equation (\ref{ea2B}) also has only the trivial identically null solution because the operator $e^{a^2\Box}$ has no zero modes.
In other words, the only solution to the homogeneous partial differential equations (\ref{N=4}) and (\ref{N=6}), as well as (\ref{N=2}) and (\ref{ea2B}) is $\phi\equiv0$.
Well, this can be actually good when it comes to solving the associated non-homogeneous partial differential equation.  If $J(x)$ is a given external source function, then the solution to
\begin{equation}\label{ea2BJ}
e^{a^2\Box}\phi = J(x)
\,,
\end{equation}
can be obtained directly as
\begin{equation}
\phi = e^{-a^2\Box}J(x)
\end{equation}
with no further freedom for boundary conditions.  Similarly, if $N$ is even, we can immediately write the unique solution to the non-homogeneous partial differential equation
\begin{equation}\label{EL2NJ}
\sum_{n=0}^N\frac{a^{2(n-1)}\Box^n}{n!}\phi=J(x)
\,,
\end{equation}
as
\begin{equation}\label{EL2NJsol}
\phi^J_{(N)}(x)={\left( \sum_{n=0}^N\frac{a^{2(n-1)}\Box^n}{n!} \right)}^{-1}J(x)\,
\,,
\end{equation}
where a proper meaning can be assigned to the RHS by means of the Fourier transform of $J(x)$.  In fact, after taking the Fourier transform of both sides of equation (\ref{EL2NJ}), we see that the RHS of (\ref{EL2NJsol}) is given in terms of the external source $J(y)$ as
\begin{equation}\label{EL2NJsol2}
\phi^J_{(N)}(x)=  \int \frac{d^Dpd^Dy}{(2\pi)^D}
e^{-ip\dot (x-y)}
{\left(
\sum_{n=0}^N{(-1)}^n\frac{a^{2(n-1)}}{n!}p^{2n}
\right)}^{-1}
J(y)\,
\,.
\end{equation}

Concerning the remaining odd values for $N$, we turn next to the standard technique of first obtaining the field propagator and then writting the general solution for (\ref{EL2NJ}) as a linear combination of contributions from elementary delta function sources.  In terms of a given external current $J(x)$, we may write the functional generator associated to action (\ref{S2N}) as
\begin{equation}
Z^{(2N)}[J]={\cal N}\int[d\phi]\exp
\Big\lbrace iS^{(2N)}+i\int d^Dx \, J(x)\phi(x)
\Big\rbrace
\,,
\end{equation}
with
\begin{equation}
{\cal N}^{-1}\equiv\int[d\phi]\exp
\Big\lbrace iS^{(2N)}
\Big\rbrace
\,.
\end{equation}
The propagator for the scalar field $\phi(x)$ can be immediately computed as
\begin{equation}\label{Prop}
D^{(2N)}=
\frac{-ia^2}{\displaystyle\sum_{n=0}^N (-1)^n\frac{{(a^2p^2)}^n}{n!}}
\end{equation}
and has a real pole for odd $N$ at $p^2=q_N/a^2$ with $q_N$ denoting the only real solution to the polynomial equation (\ref{f=0}).  Note in particular that for $N=1$, this propagator reduces to the well-known ordinary KG propagator given by
\begin{equation}
D^{(2)}=\frac{-ia^2}{1-a^2p^2}=\frac{i}{p^2-1/a^2}
\,.
\end{equation}
With this final expression for the generalized propagator we end our analyses of the classical solutions to the models (\ref{S2N}) and (\ref{Sphi})  concerning both the homogeneous and nonhomogeneous cases.

\section*{Conclusion and Final Remarks}
We have proposed a natural higher-order generalization of the KG action and investigated its energy-momentum tensor and classical solutions.  Given a natural number $N$, we have seen that action $S^{(2N)}$ in equation (\ref{S2N}) leads to the $2N$-th-order partial differential equation
\begin{equation}\tag{\ref{EL2N}}
\sum_{n=0}^N\frac{a^{2(n-1)}\Box^n}{n!}\phi=0
\,,
\end{equation}
whose solutions were obtained in the specific cases of odd and even $N$.  We have shown that for even values of $N$, the homogeneous equation (\ref{EL2N}) has only the trivial solution $\phi\equiv0$.  Also the infinite-order homogeneous equation (\ref{ea2B}) does not have non-trivial solutions for $\phi$.  On the other hand, for odd values of $N$ we have obtained the general solution of (\ref{EL2N}) in equation (\ref{gs}).  We have seen that the general solution is virtually the same as the ordinary second-order KG equation with a different value for the mass $m$ for the field excitation modes.  More precisely we have obtained a new mass value given by $m=q_N^{1/2}/a$ as can be directly inferred from equation (\ref{DR}).   Concerning the non-homogeneous case, in which an external source $J(x)$ is introduced, we have shown that for even values of $N$ and for the infinite-order case, we have a unique solution for $\phi$ without further freedom for adjusting any given boundary conditions.   On the contrary, if $N$ is odd, a solution satisfying the desired boundary conditions can be constructed from the field propagator in the usual way.   We have obtained the generalized propagator in equation (\ref{Prop}) and, in consonance with the mentioned result for the homogeneous case, shown that it is indeed very similar to the usual second-order KG propagator with a new mass $m=q_N^{1/2}/a$.  Actually, it has the advantage of possessing higher momentum powers in the denominator, thus enhancing convergence in quantum field theory perturbative calculations, without changing the degrees of freedom of the second-order model.  This final analysis shows a strong connection to the quantum field standard regularization program -- the new $m=q_N^{1/2}/a$ can thought as a renormalized mass.  A concrete realization of this idea can be seen in reference \cite{Ji:2019phv} in which the Pauli-Villars regularization method is elucidated in the context of 
a fourth-order theory for gauge fields.

\end{document}